\documentstyle[12pt]{article}
\begin{document}
\title{{\bf GENERAL RELATIVISTIC EFFECTS ON THE INDUCED ELECTRIC
FIELD EXTERIOR TO PULSARS}}
\author{\bf S. SENGUPTA \thanks{e-mail : sujan@iiap.ernet.in} \\
\normalsize {\em Indian Institute Of Astrophysics, Bangalore 560 034,
India}}
\date{}
\maketitle
\bigskip
{\bf To appear in The Astrophysical Journal, August 10, 1995}
\clearpage
\begin{abstract} \normalsize\noindent
The importance of general relativity to the induced electric
field exterior to pulsars has been investigated by assuming
aligned vacuum and non-vacuum magnetosphere models. For this
purpose the stationary and axisymmetric vector potential in
Schwarzschild geometry has been considered and the corresponding
expressions for the induced electric field due to the rotation
of the magnetic dipole have been derived for both vacuum and
non-vacuum conditions.  Due to the change in the magnetic dipole
field in curved spacetime the induced electric field also
changes its magnitude and direction and increases significantly
near the surface of the star.  As a consequence the surface
charge density, the acceleration of charged particles in vacuum
magnetospheres and the space charge density in non-vacuum
magnetosphere greatly increase near the surface of the star. The
results provide the most general feature of the important role
played by gravitation and could have several potentially
important implications for the production of high-energy
radiation from pulsars. \\ {\bf Subject headings:}
electromagnetism : theory -- pulsars : general -- relativity --
stars : neutron
\end{abstract}
\clearpage
{\bf 1. INTRODUCTION}

The realization that pulsars are most likely to be neutron stars
surrounded by a dense magnetosphere (Gold 1968; Pacini 1968) and
the pioneering work by Goldreich and Julian (1969) have provided
the basic physical model on which virtually all particular
models have been based. At present two general types of models
are being considered in-order to understand the phenomenon of
high energy radiation from pulsars viz., the polar cap model
(Arons 1983) and the outer gap model (Cheng, Ho \& Ruderman
1986). Goldreich and Julian first realized that inspite of the
intense surface gravity, the neutron star must possess a dense
magnetosphere. Unfortunately the actual role played by
gravitation in this aspect has not been investigated so far in
detail.

General relativistic effects near pulsars have been explored
only in the context of emission models for X-ray pulsars and
gamma-ray burst. The effect of light bending in X-ray burst and
X-ray pulsars was studied by Meszaros and Riffert (1988),
Riffert, Meszaros \& Bagoly (1989). Recently Gonthier and
Harding (1994) have examined the importance of general
relativistic corrections to the production of gamma rays near
the surface of neutron stars.

It is well-known that the induced electric field due to the
rapid rotation of the magnetic dipole of a neutron star plays
the main role in determining the characteristic of the pulsar
magnetosphere since this quantity determines the number density
of charged particles torn off from the surface of the star, the
acceleration of charged particles and hence the production of
electromagnetic radiation, etc. Since the magnetic field in
curved spacetime differs in both magnitude and direction from
that in flat spacetime, the induced electric field is also
different in curved spacetime. Therefore it is very important to
include general relativity in any discussion of pulsar
magnetosphere.

In this paper the importance of  general relativistic
corrections to the induced electric field external to a neutron
star has been investigated quantitatively by considering the
simplest aligned vacuum and non-vacuum magnetosphere models.

The paper is organized in the following way: In section 2 the
Schwarzschild metric and the corresponding orthonormal tetrad
components alongwith the transformation rules are given. In
section 3 we discuss the dipole magnetic field in Schwarzschild
spacetime which has been used to determine the induced electric
field in section 4.  In section 4 a detail quantitative
discussion on the effects of spacetime curvature to the various
important quantities  governed by the electric field has been
presented. For simplicity we have assumed the vector potential
to be stationary and axisymmetric. Finally we conclude the
discussions in section 5.

{\bf 2. THE SPACETIME METRIC}

Since we are interested on the electromagnetic field exterior to
the neutron star, we consider the Schwarzschild metric as it
provides a simple description of the spacetime geometry exterior
to neutron stars.

The metric is given by :
      \begin{eqnarray}
ds^{2}=(1-2m/r)c^{2}dt^{2}-(1-2m/r)^{-1}dr^{2}-r^{2}d\theta^{2}
-r^{2}\sin^{2}\theta d\phi^{2}
      \end{eqnarray}
where $m=MG/c^{2}$, M being the total mass of the star.

The non-zero components of the orthonormal tetrad
$\lambda_{(\alpha)}^{i}$ of the local Lorentz frame for
Schwarzschild geometry are given by
$$\lambda_{(t)}^{t}=(1-2m/r)^{-1/2},
\lambda_{(r)}^{r}=(1-2m/r)^{1/2},
\lambda_{(\theta)}^{\theta}=\frac{1}{r}, \lambda_{(\phi)}^{\phi}
=\frac{1}{r\sin\theta}.$$

If $F_{(\alpha \beta)}$ are the components of the
electromagnetic field tensor in local Lorentz frame, then the
components of the electromagnetic field tensor $F_{ij}$ defined
in the curved spacetime through
      \begin{eqnarray}
F_{(\alpha\beta)}=\lambda_{(\alpha)}^{i}\lambda_{(\beta)}^{j}F_{ij}
       \end{eqnarray}

{\bf 3. THE DIPOLE MAGNETIC FIELD IN CURVED SPACETIME}

Since the magnetic field of a neutron star is generally
approximated to be that of a pure magnetic dipole, in flat
spacetime we can write the components of the magnetic field (in
polar coordinates) as
     \begin{eqnarray}
{\bf B}=\frac{2\mu}{r^{3}}(\cos\theta,
\frac{1}{2}\sin\theta, 0)
      \end{eqnarray}
where $\mu$ is the magnetic dipole moment.

The magnetic dipole has to be modified in curved spacetime. In
Schwarzschild spacetime the electromagnetic potential $A_{i}$ is
given by (Petterson 1974; Wasserman \& Shapiro 1983)
$$A_{i}=(0,0,A_{\phi},0)$$
with
     \begin{eqnarray}
A_{\phi}=-\frac{3\mu\sin^{2}\theta}{8m^{3}}[r^{2}\ln(1-2m/r)
+2mr+2m^{2}]
      \end{eqnarray}

Using the definition $F_{ij}=(A_{j,i}-A_{i,j})$ we get the
components of the magnetic field in Schwarzschild geometry as
      \begin{eqnarray}
F_{r\phi}=-\frac{3\mu\sin^{2}\theta}{4m^{2}}[\frac{r}{m}\ln(1-
2m/r)+(1-2m/r)^{-1}+1]
      \end{eqnarray}
       \begin{eqnarray}
F_{\theta \phi}=-\frac{3\mu r^{2}}{4m^{3}}\sin\theta\cos\theta
[\ln(1-2m/r)+\frac{2m}{r}(1+m/r)]
       \end{eqnarray}

Therefore using equation (2) we get
      \begin{eqnarray}
B_{r}(Sch)=-\frac{3\mu\cos\theta}{4m^{3}}[\ln(1-2m/r)
+\frac{2m}{r}(1+m/r)]
       \end{eqnarray}
       \begin{eqnarray}
B_{\theta}(Sch)=\frac{3\mu\sin\theta}{4m^{2}r}
[\frac{r}{m}\ln(1-2m/r)+(1-2m/r)^{-1}+1](1-2m/r)^{1/2}
       \end{eqnarray}

It should be mentioned that the above expressions are
specialized to the lowest order expansion in terms of spherical
harmonics. Nevertheless they are valid in the context of neutron
stars since the magnetic field is not so strong relative to the
gravitational field such that the spacetime geometry could be
affected significantly by the magnetic field.

{\bf 4. THE INDUCED ELECTRIC FIELD IN CURVED SPACETIME}

In order to investigate the effect of curved spacetime on the
electric field exterior to a rotating neutron star, we consider
two simplest models, viz., (1) the aligned vacuum rotator and
(2) the aligned non-vacuum rotator. Since the effect of
curvature would be the same, the results will provide the most
general feature for any magnatospheric model which assumes
stationary $(\frac{\partial}{\partial t}=0)$ and axisymmetric
$(\frac{\partial}{\partial \phi}=0)$ behavior of the vector
potential. In the present study we have ignored any effects due
to magnetosphere plasma. Also we have neglected the dragging of
inertial frame due to rotation which is insignificant at the
exterior of even millisecond pulsars.

{\bf 4.1. The Aligned Vacuum Rotator}

Since the derivations are nothing but a generalization of
Goldreich - Julian solution, we briefly discuss it before going
to derive the electromagnetic field in curved spacetime although
it can be found in any text book on Pulsars e.g., Shapiro and
Teukolsky (1983).

First of all it is considered that a rotating neutron star has
an aligned dipole external magnetic field given by equation (3).
The stellar material is assumed to be an excellent conductor so
that just inside the star an electric field will be present
which satisfies
       \begin{eqnarray}
{\bf E}^{int} + \frac{{\bf \omega}\times{\bf r}}{c}\times {\bf B}^{int} =0
       \end{eqnarray}
where ${\bf \omega}$ is the angular velocity vector of the star.

Here and afterwards the superscript `int' is to be understood to
represent the quantities just inside the star and not far bellow
the surface where the above condition may not be valid.

Assuming there are no surface currents, both the normal and the
tangential components of {\bf B} are continuous across the
stellar surface. Thus just inside the surface the magnetic field
can be written by taking $r=R$ in equation (3) where $R$ is the
radius of the star. Then equation (9) gives for {\bf E} just
inside the surface
     \begin{eqnarray}
 {\bf E}^{int} =\frac{2\mu\omega}{cR^{2}}(\frac{1}{2}\sin^{2}\theta,
-\sin\theta\cos\theta, 0)
     \end{eqnarray}
The tangential component of {\bf E} is continuous across the surface,
so just outside the
star equation (10) implies
     \begin{eqnarray}
E_{\theta}^{ext} =-\frac{\partial}{\partial\theta}
(\frac{\mu\omega\sin^{2}\theta}{cR^{2}})=\frac{\partial}{\partial\theta}
[\frac{2\mu\omega}{3cR^{2}}P_{2}(\cos\theta)]
     \end{eqnarray}
where $P_{2}$ is the Legendre polynomial of second degree.

In the present case since the exterior is assumed to be vacuum, so
$${\bf E}^{ext}=-\nabla\phi$$
where
     \begin{eqnarray}
\nabla^{2}\phi=0
    \end{eqnarray}

In order to satisfy the boundary condition given by equation
(11) at $r=R$, the solution of equation (12) must be
$$\phi=-\frac{2\mu\omega R^{2}}{3cr^{3}}P_{2}(\cos\theta)$$
and hence
     \begin{eqnarray}
{\bf E}^{ext}= -\frac{\mu\omega R^{2}}{cr^{4}}[(3\cos^{2}\theta-1),
2\sin\theta\cos\theta,0]
     \end{eqnarray}

Now we will derive the electric components exterior to a
rotating neutron star for Schwarzschild background geometry with
the basic assumptions taken in the above discussions unaltered.

Equation (9) can be written in its covarient form as
     \begin{eqnarray}
F^{\beta}_{\delta}u^{\delta}=0
    \end{eqnarray}
where $u^{\delta}$ are the components of the four velocity vector.

{}From equation (14) we obtain
      \begin{eqnarray}
F_{rt}=F_{\phi r}\frac{u^{\phi}}{u^{t}}
      \end{eqnarray}
      \begin{eqnarray}
F_{\theta t}=F_{\phi \theta}\frac{u^{\phi}}{u^{t}}
      \end{eqnarray}
(Note that $u=(u^{t},0,0,u^{\phi})$ for the present case.)

Using equations (5) and (6) and noting that in Schwarzschild geometry
$$E_{r}=F_{(rt)}=F_{rt}$$
$$ E_{\theta}=F_{(\theta t)}=F_{\theta t}/[r(1-2m/r)^{1/2}]$$
and taking
$$\frac{u^{\phi}}{u^{t}}=\frac{\omega}{c}$$
we obtain the induced electric field components just inside the
surface \\ $(r\simeq R)$ as
      \begin{eqnarray}
E_{r}^{int}=\frac{3\mu\omega\sin^{2}\theta}{4m^{2}c}[(1-2m/r)^{-1}+
\frac{r}{m}\ln(1-2m/r)+1]
      \end{eqnarray}
      \begin{eqnarray}
E_{\theta}^{int}=\frac{3\mu\omega r\cos\theta\sin\theta}{4cm^{3}}
[\ln(1-2m/r)+\frac{2m}{r}(1+\frac{m}{r})](1-\frac{2m}{r})^{-1/2}
      \end{eqnarray}

If $f(r)$ and $g(r)$ are two arbitrary functions which solely
determine the spacetime curvature effect then in curved
spacetime we can write
      \begin{eqnarray}
E_{r}^{int}(curved)=\frac{\mu\omega}{cr^{2}}\sin^{2}\theta f(r)
       \end{eqnarray}
       \begin{eqnarray}
E_{\theta}^{int}(curved)=-\frac{2\mu\omega}{cr^{2}}\sin\theta
\cos\theta g(r)
       \end{eqnarray}

Comparing equations (19) and (20) with equations (17) and (18)
we obtain
     \begin{eqnarray}
f(r)=\frac{3r^{2}}{4m^{2}}[(1-2m/r)^{-1}+\frac{r}{m}\ln(1-2m/r)+1]
      \end{eqnarray}
      \begin{eqnarray}
g(r)=-\frac{3r^{3}}{8m^{3}}[\ln(1-2m/r)+\frac{2m}{r}(1+\frac{m}{r})
(1-\frac{2m}{r})^{-1/2}
       \end{eqnarray}

In the above derivations we have assumed that the metric can
well describe the spacetime curvature just bellow the surface
$(r\simeq R)$ of the star, i.e., the curvature effect just
bellow the surface is the same as that at the surface to the
star.

In flat spacetime the components of the exterior electric field
is given by equation (13).

Therefore in Schwarzschild geometry the components of the
exterior electric field can be written as
       \begin{eqnarray}
E_{r}^{ext}(Sch) &=&-\frac{\mu\omega R^{2}}{cr^{4}}(3\cos^{2}\theta-
1)f(r) \nonumber \\
&=& -\frac{3\mu\omega R^{2}}{4cm^{2}r^{2}}(3\cos^{2}\theta-1)
[(1-2m/r)^{-1}+\frac{r}{m}\ln(1-2m/r)+1]
      \end{eqnarray}
       \begin{eqnarray}
E_{\theta}^{ext}(Sch) &=&-\frac{2\mu\omega R^{2}}{cr^{4}}
\sin\theta\cos\theta g(r) \nonumber \\
&=& \frac{3\mu\omega R^{2}\cos\theta\sin\theta}
{4cm^{3}r}[\ln(1-2m/r)+\frac{2m}{r}(1+\frac{m}{r})]
(1-\frac{2m}{r})^{-1/2}
     \end{eqnarray}

Clearly ${\bf E^{ext}(curved)\rightarrow E^{ext}(flat)}$ as
$r\rightarrow\infty$ as desired.

Taking as examples, the Crab pulsar with mass M equal to 1.4
solar mass, radius R equal to $10^{6} cm$, period P equal to 33
ms and the magnetic dipole moment $\mu$ equal to $2\times10^{30}
Gcm^{3}$ we show in Figure 1 the $r$ component of the quadrapole
electric field at an angle $\theta=0^o$ to the axis of rotation
as a function of distance from the stellar surface for flat and
curved spacetimes. It is found that a significant increase in
the electric field intensity near the surface is caused due to
the inclusion of general relativistic effect.

Now we have the functional form for the dipole magnetic field
and the quadrapole electric field in flat spacetime as well as
in Schwarzschild spacetime.  We would like to see how the field
lines differ in two different geometries.  The field lines
represent a constant magnetic and electric flux $b(\theta)$ and
$e(\theta)$ respectively and can be found by rotating a given
field line about the z-axis in a spherical co-ordinate system
with the magnetic dipole aligned with the z-axis. The fluxes
from $\theta=0$ to $\theta$ are given by
        \begin{eqnarray}
b(\theta)=\int_{0}^{\theta} B_{r}.d{\bf a}=\int^{\theta}_{0}
B_{r}r^{2}d\Omega
       \end{eqnarray}
       \begin{eqnarray}
e(\theta)=\int_{0}^{\theta}
E_{r}.d{\bf a}=\int^{\theta}_{0} E_{r}r^{2}d\Omega
       \end{eqnarray}

The magnetic field lines in flat spacetime and in curved
spacetime have been presented by Gonthier and Harding (1994).

For an induced quadrapole electric field in flat spacetime
the field lines are
     \begin{eqnarray}
e(\theta)=-\frac{2\pi\mu\omega R^{2}}{cr^{2}}\sin^{2}\theta
\cos\theta=constant
      \end{eqnarray}

In curved spacetime the field lines are given by
        \begin{eqnarray}
e(\theta) &=& -\frac{3\pi\mu\omega R^{2}}{2m^{2}c}\sin^{2}
\theta\cos\theta[(1-2m/r)^{-1}+\frac{r}{m}\ln(1-2m/r)+1] \nonumber \\
&=& constant
        \end{eqnarray}

Taking the Crab pulsar as an example with a period of 33 ms we
show in Figure 2a the quadrapole electric field lines in the z-x
plane out to 40 stellar radii, beyond which the effect of
curvature becomes insignificant. Figure 2b shows a closer view
of the same field lines.

Since the normal component of the electric field is
discontinuous for the case of aligned vacuum rotator, the
corresponding surface charge density can be written in flat
spacetime as
       \begin{eqnarray}
\rho=\frac{1}{4\pi}(E_{r}^{ext}-E_{r}^{int})=-\frac{\mu\omega}{2\pi
cR^{2}}\cos^{2}\theta
       \end{eqnarray}

In curved spacetime the same quantity can be written as
     \begin{eqnarray}
\rho=-\frac{3\mu\omega}{8\pi cm^{2}}\cos^{2}\theta[(1-2m/R)^{-1}
+\frac{R}{m}\ln(1-2m/R)+1]
     \end{eqnarray}

Since the effect of curvature increases the electric field
intensity near the surface, therefore in curved spacetime the
surface charge density is much higher than that in flat
spacetime.  For a pulsar with mass $1.4 M_{\odot}$ and radius 10
km the surface charge density at the pole in curved spacetime is
twice of that in flat spacetime.

The Lorentz invariant scalar product ${\bf E^{ext}.B}$ does not
vanish. This quantity gives a measure of the force which a
co-rotating charged particle feels in the direction of the
magnetic field.

In flat spacetime
     \begin{eqnarray}
{\bf E^{ext}.B}=-\frac{4\mu^{2}\omega R^{2}\cos^{3}\theta}{cr^{7}}
    \end{eqnarray}
leading to an acceleration of a charged particle in the
direction of the magnetic field:
      \begin{eqnarray}
a=\frac{e{\bf E^{ext}.B}}{m_{p}|{\bf B}|}=-\frac{4e}{m_{p}c}
\frac{\mu\omega R^{2}}{r^{4}}\cos^{3}\theta(3\cos^{2}\theta+1)^{-1/2}
    \end{eqnarray}
where $m_{p}$ is the mass of the charged particle and $e$ is
its charge. The corresponding quantities in curved spacetime
can be obtained by using equations
(7), (8) and (24) and are given by
      \begin{eqnarray}
{\bf E^{ext}.B}=\frac{9\mu^{2}\omega R^{2}}{8m^{5}cr^{2}}F(r)
G(r)\cos^{3}\theta
      \end{eqnarray}
and
     \begin{eqnarray}
a=\frac{3e}{2m_{p}c}\frac{\mu\omega R^{2}}{m^{3}r^{2}}F(r)G(r)
\cos^{3}\theta [\frac{\cos^{2}\theta}{m^{2}}G^{2}(r)+
\frac{\sin^{2}\theta}{r^{2}}(1-\frac{2m}{r})F^{2}(r)]^{-1/2}
     \end{eqnarray}
where
    \begin{eqnarray}
F(r)=[(1-2m/r)^{-1}+\frac{r}{m}\ln(1-2m/r)+1]
    \end{eqnarray}
      \begin{eqnarray}
G(r)=\ln(1-2m/r)+\frac{2m}{r}(1+\frac{m}{r})
     \end{eqnarray}

In Figure 3 we present the variation of acceleration of charged
particles with distance by considering the Crab pulsar with a
rotational period 33ms.  The acceleration has been measured at
the angle $\theta=0^{o}$ to the axis of rotation.  Since
gravitation greatly increases the electric field intensity near
the surface of the star, the acceleration of charged particles
is much higher in curved spacetime than that in flat spacetime
near the surface. This result indicates that the radiation
spectrum of pulsars may significantly be influenced by the
intense gravitational field of neutron stars and  requires
further attention.

Since in realistic situation the magnetosphere must be filled up
with charge plasma, therefore the various quantities discussed
in this section may not represent the actual situation. However
the effect of curvature remains the same whether the region
under consideration is vacuum or not, since the gravitational
effect due to the magnetospheric plasma is negligible.

{\bf 4.2. The Aligned Non-vacuum Rotator}

It is now well accepted that the magnetosphere of neutron stars
must be filled up with plasma such that ${\bf E.B}=0$ everywhere
within the light cylinder.  In the Goldreich-Julian model the
source of plasma is assumed to be free field emission from the
surface.

In this section we will briefly discuss the effect of spacetime
curvature on the induced electric field by considering the
aligned non-vacuum rotator model.

In order to investigate the effect of curvature to the induced
electric field we consider the Godreich-Julian solution which
simply corresponds to extending the interior solution to
infinity with space charge density continuing smoothly through
the surface. We neglect, for simplicity any effects due to the
magnetospheric plasma. However inclusion of the same would not
change the general feature of the curvature effect.

Therefore, for this case, equations (17) and (18) express the
components of the electric field outside the star in
Schwarzschild spacetime with $r$ being extended from just inside
the surface to infinity (the superscript `int' should now be
ignored).

The corresponding expression in flat spacetime can be written as
     \begin{eqnarray}
{\bf E}(flat) = \frac{\mu\omega}{cr^{2}}(\sin^{2}\theta,
-2\sin\theta\cos\theta, 0)
     \end{eqnarray}

Asymptotically as $r\rightarrow\infty, {\bf E}(curved)\rightarrow{\bf E}
(flat)$ as it must be the boundary condition.

{}From equation (10) we obtain the electric field lines in flat
spacetime as
       \begin{eqnarray}
e(\theta)=\frac{2\pi\mu\omega}{3c}[\cos^{3}\theta-3\cos\theta
+2]=constant
        \end{eqnarray}
{}From equation (17) we obtain the same in curved spacetime as
      \begin{eqnarray}
e(\theta)&=& \frac{\pi\mu\omega r^{2}}{2cm^{2}}[\cos^{3}\theta-
3\cos\theta+2][(1-2m/r)^{-1}+\frac{r}{m}\ln(1-2m/r)+1] \nonumber \\
&=& constant
      \end{eqnarray}

Clearly the electric field lines for flat spacetime is
independent of $r$ whereas those in Schwarzschild background
geometry depend on $r$ only near the surface.  At comparatively
large distance the field lines for curved spacetime
asymptotically become independent of distance and hence it is
not possible to plot them at a large distance. Also, since the
electric field is continuous from the interior to the exterior
of the star for the non-vacuum case, the expressions are
independent of the radius of the star.

Taking, as earlier, the Crab pulsar with a period of 33ms we
show in Figure 4 the induced electric field lines in curved
spacetime in the x-z plane out to 4 stellar radii. Beyond this
distance the field lines almost become independent of r and it
is not possible to present them graphically.

Figure 5 shows the complete picture of the non-vacuum
magnetosphere in curved spacetime with both the induced electric
field lines and the dipole magnetic field lines obtained by
using the expression for the magnetic field lines:
     \begin{eqnarray}
b(\theta)=-\frac{3\pi\mu r^{2}\sin^{2}\theta}{4m^{3}}[\ln(1-2m/r)+
\frac{2m}{r}(1+\frac{m}{r})]=constant
     \end{eqnarray}

In both the plots we have taken, for convenience, the electric
and the magnetic field lines from the interior of the star and
extended it to the exterior.  However, far bellow the surface
the spacetime metric considered here, cannot describe the
geometry of the region, hence in the plots they are not shown at
a distance far bellow the surface of the star.

{}From equations (27) and (28) we notice that the ratio between
the electric fluxes in curved and flat spacetime for the vacuum
model is the same to that for the non-vacuum case and is given
by
      \begin{eqnarray}
\frac{e(\theta)[curved]_{vac}}{e(\theta)[flat]_{vac}}=
\frac{e(\theta)[curved]_{non-vac}}{e(\theta)[flat]_{non-vac}}=
\frac{3r^{2}F(r)}{4m^{2}}
      \end{eqnarray}
where $F(r)$ is given by equation (35).

Hence the effect of spacetime curvature to the electric field lines
remains the same in this case as well.

In flat spacetime the space charge density is given by
       \begin{eqnarray}
\sigma=\frac{1}{4\pi}div{\bf E}=-\frac{\mu\omega}{2\pi cr^{3}}
(3\cos^{2}\theta-1)
        \end{eqnarray}
which is always quadrapolar.

The corresponding expression in curved spacetime can be written as:
         \begin{eqnarray}
\sigma &=&\frac{3\mu\omega}{16\pi cm^{2}}\{[\frac{4\sin^{2}\theta}{r}
(1-2m/r)^{-1}+\frac{3\sin^{2}\theta}{m}\ln(1-2m/r)-\frac{2m\sin^{2}
\theta}{r^{2}}(1-2m/r)^{-2}+ \nonumber \\
& & \frac{2\sin^{2}\theta}{r}]+\frac{(3\cos^{2}\theta-1)}{m}
[\ln(1-2m/r)+\frac{2m}{r}(1+\frac{m}{r})](1-\frac{2m}{r})^{-1/2}\}
     \end{eqnarray}

Figure 6 presents the space charge density as a function of
distance from the surface of the Crab pulsar at an angle
$\theta=0^{o}$ to the axis of rotation.  In all the figures the
various quantities have been measured in CGS units and their
absolute values have been taken in logarithmic scale unless
mentioned otherwise.

Since  gravitation increases the electric field intensity at the
surface, more particles get torn off the surface. The electric
field required to pull ions from the molecular chains with the
ions distributed in a one dimensional lattice along the chain
and with an outer sheath of electrons, is of the order
$10^{12}Vcm^{-1}$ (Ruderman \& Sutherland 1975). Therefore the
effect of gravitation in some pulsars may cause the surface
electric field to reach this cut off value and hence may enable
to pull ions. However, ions may be torn off from the surface due
to other effects (Mitchel 1982). In that case, the inclusion of
gravitation would increase the number density of ions, in the
same way it increases the number density of electrons.

{\bf 5. CONCLUSIONS}

We have presented a quantitative discussion on the effects of
spacetime curvature to the induced electric field of a pulsar.
Considering stationary and axisymmetric electromagnetic vector
potential in Schwarzschild background geometry the components of
the induced electric field that arises due to the rotation of
the magnetic dipole, have been derived and compared with that in
flat spacetime. It is found that spacetime curvature increases
the electric field intensity significantly near the surface of
the star. As a consequence, in vacuum magnetosphere the
acceleration of charged particles along the direction of the
magnetic field increases significantly near the surface. Also
the increase in the electric field intensity near the surface
results in increasing the surface charge density significantly.
The quadrapole electric field lines have been presented for Crab
pulsar, as an example, in both curved and flat spacetime and it
is found that the difference in the field lines for the curved
and the flat spcetimes is well distinguishable upto 40 stellar
radii.

In a plasma filled magnetosphere the electric field lines are
independent of distance in flat spacetime but in curved
spacetime they depend on $r$ near the surface. However, the field
lines in curved spacetime asymptotically become independent of $r$
at large distance.  The induced electric field lines in curved
spacetime have been presented by taking Crab pulsar, as an
example and the combined dipole magnetic field lines and the
induced electric field lines near the surface of the star for
curved spacetime have been presented as well.

Due to the increase in the electric field intensity near the
surface the space charge density of a plasma filled
magnetosphere increases significantly from that in flat
spacetime. All the quantities measured in curved spacetime,
however asymptotically reach their respective values in  flat
spacetime as a result of the usual boundary condition. Since the
ratio of the components of the electric field in curved
spacetime to that in flat spacetime is the same, therefore the
effects of gravitation on various electromagnetic mechanisms
remain the same for all types of magnetospheric models which
assume stationary and axisymmetric vector potentials. Hence the
present results are very important in the context of radiation
mechanisms in pulsars.

Though we have considered the most idealized models of the
pulsar magnetosphere the effects of spacetime curvature to the
induced electric field would remain the same provided there is
no dipole or gravitational radiation.  The effect of inertial
frame dragging is negligible even for millisecond pulsars but
the depurture from spherical symmetry due to rapid rotation of
the star could alter the scenario significantly. Nevertheless,
the present results would provide reasonably well insight on the
effect of spacetime curvature to the induced electric field
outside a rotating neutron star with its magnetic axis aligned
to the rotational axis.

{\bf ACKNOWLEDGEMENTS}

I am grateful to the referee for many valuable comments and
suggestions which have not only improved the clarity of the
paper in great extent but also have removed some serious errors
in an early version of the manuscript.  Thanks are due to B. R.
Iyer for discussions.

\clearpage
\begin{center}
{\bf REFERENCES}
\end{center}
\vskip1cm
\begin{flushleft}
Arons, J. 1983, ApJ, {\bf 266}, 215 \\ Cheng, K. S., Ho, C. \&
Ruderman, M. A. 1986, ApJ, {\bf 300}, 495 \\ Gold, T. 1968,
Nature, {\bf 218}, 731 \\ Goldreich, P. \& Julian, W. H. 1969,
ApJ, {\bf 157}, 869 \\ Gonthier, P. L. \& Harding, A. K. 1994,
ApJ, {\bf 425}, 767 \\ Meszaros, P. \& Riffert, H. 1988, ApJ,
{\bf 327}, 712 \\ Michel, F. C. 1982, Rev. Mod. Phys. {\bf 54},
1 \\ Pacini, F. 1968, Nature, {\bf 219}, 145 \\ Petterson, J. A.
1974, Phys. Rev. {\bf D10}, 3166 \\ Riffert, H., Meszaros, P. \&
Bagoly, Z. 1989, ApJ, {\bf 340}, 443 \\ Ruderman, M. A. \&
Sutherland, P. G. 1975, ApJ, {\bf 196}, 51 \\ Shapiro, S. L. \&
Teukolsky, S. A. 1983, Black Holes, White Dwarfs and Neutron
Stars, John Wiley (New York)\\ Wasserman, I. \& Shapiro, S. L.
1983, ApJ, {\bf 265}, 1036
\end{flushleft}
\clearpage
\begin{center}
{\bf FIGURE CAPTIONS}
\end{center}

{\bf FIG 1. --} The r component of the quadrapole electric field
at an angle $\theta=0^o$ to the axis of rotation for Crab pulsar
as a function of distance from the stellar surface. Solid curve
represents the field intensity for curved spacetime while dashed
curve for flat spacetime.

{\bf FIG 2a. --} Quadrapole electric field lines (induced) for
the Crab pulsar with $\mu=2\times 10^{30}Gcm^{3}$. Solid curves
represent field lines for curved spacetime while dashed curves
for flat spacetime. The corresponding value assigned to the
field line is the same for both curved and flat spacetimes.

{\bf FIG 2b. --} A closer view of the quadrapole electric field
lines shown in Figure 2a.

{\bf FIG 3. --} The acceleration of an electron in the direction
of the magnetic field at an angle $\theta=0^o$ to the axis of
rotation for Crab pulsar as a function of distance from the
stellar surface. Solid curve -- curved spacetime; dashed curve
-- flat spacetime.

{\bf FIG 4. --} The induced electric field lines in curved
spacetime for the Crab pulsar with a period 33ms and
$\mu=2\times10^{30}Gcm^{3}$ by considering the magnetosphere to
be plasma filled (see text).

{\bf FIG 5. --} The combined dipole magnetic and induced
electric field lines in curved spacetime for the Crab pulsar
with $\mu=2\times10^{30}Gcm^{3}$ by considering plasma filled
magnetosphere. Solid curves represent the induced electric field
lines while dashed curves for the dipole magnetic filed lines.

{\bf FIG 6. --} The space charge density at an angle
$\theta=0^{o}$ to the axis of rotation for the Crab pulsar as a
function of distance from the stellar surface. Solid curve --
curved spacetime; dashed curve -- flat spacetime.
\end{document}